\documentclass[onecollarge]{svjour2}       
\smartqed  
\usepackage{graphicx}
\usepackage{psfrag}
\usepackage{subfigure}

\newcommand{\be}{\begin{eqnarray}}
\newcommand{\ee}{\end{eqnarray}}
\newcommand{\bn}{\begin{eqnarray*}}
\newcommand{\en}{\end{eqnarray*}}

\newcommand{\nn}{\nonumber \\}
\newcommand{\nl}{\\}

\renewcommand{\vec}[1]{\mbox{\boldmath$#1$}}
\renewcommand{\d}{\mbox{\rm d}}
\newcommand{\gslash}[1]{\mbox{\slash{\hspace{-2mm}}$#1$}}

\newcommand{\al}{\ensuremath{\alpha}}
\newcommand{\bt}{\ensuremath{\beta}}
\newcommand{\sg}{\ensuremath{\sigma}}
\newcommand{\gm}{\ensuremath{\gamma}}
\newcommand{\dl}{\ensuremath{\delta}}
\newcommand{\lm}{\ensuremath{\lambda}}

\newcommand{\gfive}{\ensuremath{\gm^5}}

\newcommand{\Gm}{\ensuremath{\Gamma}}

\newcommand{\ze}{\ensuremath{\hat{0}}}

\newcommand{\pvec}{\ensuremath{\vec{p}}}
\newcommand{\Pvec}{\ensuremath{\vec{P}}}
\newcommand{\nvec}{\ensuremath{\vec{n}}}

\newcommand{\Rvec}{\ensuremath{\vec{R}}}

\newcommand{\sgvec}{\ensuremath{\vec{\sg}}}

\newcommand{\Dvec}{\ensuremath{\vec{D}}}
\newcommand{\Jvec}{\ensuremath{\vec{J}}}
\newcommand{\Lvec}{\ensuremath{\vec{L}}}

\newcommand{\Wvec}{\ensuremath{\vec{W}}}
\newcommand{\Gmvec}{\ensuremath{\vec{\Gm}}}
\newcommand{\Omvec}{\ensuremath{\vec{\Omega}}}

\newcommand{\nabvec}{\ensuremath{\vec{\nabla}}}

\newcommand{\hb}{\ensuremath{\hbar}}

\newcommand{\pt}[1]{\ensuremath{{\partial \over \partial #1}}}

\newcommand{\lt}{\ensuremath{\left}}
\newcommand{\rt}{\ensuremath{\right}}

\renewcommand{\d}{\mbox{\rm d}}

\begin{document}

\title{Is There an Observable Limit to Lorentz Invariance at the Compton Wavelength Scale?
}
\subtitle{Essay written for the Gravity Research Foundation 2009 Awards for Essays on
 Gravitation}


\author{Dinesh Singh         \and
        Nader Mobed 
}


\institute{D. Singh \at
              Department of Physics \\ University of Regina \\ Regina, Saskatchewan \\ S4S 0A2, Canada \\ \\
              Tel.: +1-306-585-4681\\
              Fax: +1-306-585-5659\\ \\
              \email{dinesh.singh@uregina.ca}            
           \and
           N. Mobed \at
              Department of Physics \\ University of Regina \\ Regina, Saskatchewan \\ S4S 0A2, Canada \\ \\
              Tel.: +1-306-585-4359\\
              Fax: +1-306-585-5659\\ \\
              \email{nader.mobed@uregina.ca} }

\date{March 26, 2009}

\maketitle

\begin{abstract}
The possibility of a frame-induced violation of Lorentz invariance due to non-inertial spin-1/2 particle motion
is explored in detail for muon decay while in orbit near the event horizon of a microscopic Kerr black hole.
It is explicitly shown that kinematic and curvature contributions to the muon's decay spectrum---in the absence
of any unforeseen processes due to quantum gravity---lead to its stabilization at the muon's Compton wavelength scale.
This example is emblematic of the search for unambiguous indicators to critically assess current and future
approaches to quantum gravity research.
\end{abstract}

\section{Introduction}
\label{intro}

Despite several decades of concerted effort, finding an acceptable theory of quantum gravity remains an elusive goal.
While string theory and loop quantum gravity---only two of many distinct approaches available \cite{Smolin}---provide interesting but
conflicting theoretical perspectives on the subject, it is the complete lack of reliable observational evidence that
remains the primary source of the problem.
Instead, there is a strong reliance on mathematical and philosophical assumptions to provide direction,
of which the following are commonly accepted:
\begin{description}
\item[1.]{Lorentz invariance symmetry is satisfied for all length scales involving quantum phenomena,
including the Planck length $l_P = \sqrt{\hbar G/c^3} \approx 1.616 \times 10^{-33}$~cm; and}
\item[2.]{Observations of quantum gravity {\em only} occur at the Planck scale.}
\end{description}

Given the lack of available data, it seems useful to take a more indirect approach towards solving the ``quantum gravity problem.''
For instance, consider the fact that the intrinsic properties of elementary particles are categorized in accordance
with the Poincar\'{e} symmetry group, a well-defined feature of quantum field theory in {\em flat space-time}.
However, when generalizing to quantum field theory in {\em curved space-time},
it is understood that the particle concept loses meaning, due to the lack of a unique vacuum state that is heavily
reliant on Poincar\'{e} symmetry \cite{Birrell}.
Since there are broad implications for adopting this viewpoint, it is very important to ascertain whether the
implicit assumptions built into this approach are well-founded.

To address this question, a recent investigation \cite{Singh-Mobed1} critically examines the applicability
of the Poincar\'{e} group to single-particle state quantum systems while in non-inertial motion.
This approach specifically involves the Pauli-Lubanski four-vector $\Wvec$ for spin-1/2 particles in flat space-time,
but with the canonical momentum generators $\Pvec$ in curvilinear co-ordinate form to best reflect the symmetries
of an elementary particle's classical worldline.
A surprising consequence is that the scalar operator $\Wvec \cdot \Wvec$, normally identified as a
Casimir invariant for particle spin, is actually {\em frame-dependent} due to an additive term coupling
the Pauli spin operator $\sgvec$ with a Hermitian three-vector called the
{\em non-inertial dipole operator} \cite{Singh-Mobed1,Singh1,Singh2},
in the form $\Rvec^{\hat{\imath}} = \lt(i \over 2\hbar \rt) \epsilon^{\hat{\imath}\hat{\jmath}\hat{k}}
[\Pvec_{\hat{\jmath}}, \Pvec_{\hat{k}}]$.
If $r$ is defined as the local radius of curvature for the particle's trajectory with respect to a fixed
laboratory frame \cite{Singh-Mobed1}, it is shown that $\Rvec \sim |\Pvec|/r$.
Assuming fixed momentum, it is clear that $\Rvec \rightarrow \vec{0}$ as $r \rightarrow \infty$,
while $\Rvec = \vec{0}$ identically for strictly inertial motion expressed via local Cartesian co-ordinates.
Furthermore, when $\Rvec$ is applied to studying muon decay in a circular storage ring \cite{Singh-Mobed1},
it is shown that non-inertial motion can stabilize the muon indefinitely when $r$ approaches the muon's
Compton wavelength scale.

If it is shown that a similar prediction occurs for a computation performed in a curved space-time background \cite{Singh-Mobed2},
then this is a remarkable statement that provides an important clue about the intersection between
quantum mechanics and gravitation at a length scale over {\em twenty orders of magnitude}
larger than the Planck scale, whose consequences have to be taken seriously.
It is important to note, however, that ample literature exists suggesting that quantum gravity
signatures due to energy-dependent dispersion effects for the Maxwell equations in vacuum \cite{Amelino-Camelia1,Amelino-Camelia2,Gambini}
can theoretically appear, suggesting the identification of non-trivial structure in space-time.
Furthermore, a recent review of quantum gravity phenomenology \cite{Amelino-Camelia3} shows that much
of the emphasis on searches for signatures revolves around tests of Poincar\'{e} invariance breakdown,
though not of the kind suggested in this paper and elsewhere \cite{Singh-Mobed1,Singh-Mobed2}.

The purpose of this paper is to explore modifications of the muon decay rate while in circular motion
around a microscopic Kerr black hole described in terms of Fermi normal co-ordinates,
using the spin-1/2 particle Pauli-Lubanski vector in curvilinear co-ordinates where the frame-based
contribution due to $\Rvec$ appears.
It is shown that the presence of space-time curvature {\em enhances} the stabilization of the muon,
particularly when the muon's orbital radius approaches the Compton wavelength scale, with interesting
consequences to follow.
For this paper, geometric units of $G = c = 1$ are assumed, where the curvature tensors follow the
symmetry conventions of Misner, Thorne, and Wheeler \cite{MTW}, but with $-2$ metric signature.
As well, the flat space-time gamma matrices follow the conventions given by Itzykson and Zuber \cite{Itzykson}.

\section{Review of the Hypothesis of Locality for Quantum Mechanics}
\label{sec2}

To better appreciate the conceptual subtleties that arise from considerations given in this paper,
it is useful to offer a brief review of the {\em hypothesis of locality} that is implicitly accepted
within current treatments of modern physics \cite{Singh-Mobed2,Mashhoon}.
The validity of this hypothesis may come into question when applied to quantum mechanical phenomena
under extreme conditions, particularly where high acceleration and/or gravitational effects may contribute heavily towards
a given quantum particle's propagation in space-time.

\subsection{Formalism}

The basic concept behind this hypothesis \cite{Mashhoon} is that an observer comoving with
an accelerated object corresponds one-to-one with a continuous set of instantaneously
inertial observers fixed at each moment of proper time defined along the object's worldline.
In so doing, there exists an effective {\em acceleration length} $l = a^{-1}$ that determines the range of
applicability for making spatial measurements within the locally inertial frame, where $a$ is the object's
acceleration defined with respect to some static laboratory frame.

\subsection{General Conceptual Issues with the Hypothesis of Locality}

Certainly, when considering {\em classical objects} of mass $m$ with Compton wavelength $\lm = h/m \rightarrow 0$ and/or
{\em slowly accelerated objects}, it follows that $\lm \ll l$ and no ambiguities exist when applying the
hypothesis of locality.
However, the situation becomes much more difficult to ascertain when considering {\em quantum mechanical objects}
with sufficiently long Compton wavelengths and/or {\em highly accelerated objects}, such that $\lm \sim l$.
In this instance, there now exists a ``tension'' between two fundamentally different notions of length scales,
motivated by conceptually divergent representations of what it means to identify {\em localization in space-time}.
Mathematically, a classical treatment of space-time requires the existence of a $C^\infty$ Hausdorff differentiable
manifold structure, where the concept of space-time curvature is applicable down to a
{\em mathematical point}, which is by definition a {\em structureless} feature on the manifold.
Clearly, it becomes deeply problematic to apply this limiting procedure to arbitrary precision,
given that the Heisenberg uncertainty relations must also apply \cite{Singh-Mobed1}, leading to
{\em quantum interference effects} that have significant bearing on the quantum particle's overall
physical behaviour.
This issue also becomes relevant when attempting to properly identify the location of the quantum particle's
trajectory from one moment to the next, while still incorporating quantum fluctuations about its
classical trajectory.

Furthermore, the whole notion of time itself (proper time or otherwise) becomes ambiguous within this context,
especially when the standard treatment of quantum mechanics in the {\em absence} of space-time curvature \cite{Sakurai}
requires that time is an intrinsically {\em classical parameter} associated with the space-time manifold,
while space is described in terms of a {\em quantum operator}
acting on an infinite-dimensional Hilbert space to generate position {\em observables}.
The fact that such an identifiable distinction exists between space and time measurements within quantum mechanics,
thereby going against the spirit of Lorentz covariance in general relativity, is a definite indication
of challenges within the existing formalism that need to be better understood.

\subsection{Conceptual Issues Pertaining to Quantum Gravity}

Because of these conceptual issues as it pertains to the hypothesis of locality, it is worthwhile to
consider their relevance in terms of a bona fide theory of {\em quantum gravity}, as opposed to merely
examining the effects of {\em quantum matter} propagating on a {\em classical space-time background},
in the absence of any obvious curvature-induced backreaction effects.
Unlike the ``top-down'' approaches suggested by string theory and loop quantum gravity, this paper is
motivated exclusively by ``bottom-up'' considerations, even to the point of taking a cautious attitude
about quantum field theory in curved space-time, as described in established textbook treatments
of the subject \cite{Birrell}.
With this in mind, it seems evident that the considerations to follow in this paper are highly relevant
for addressing the ``quantum gravity problem,'' specifically because it is far from obvious how
anticipated deviations from general relativity are to become manifest at sufficiently small length scales.
For example, if Wheeler's concept of {\em space-time foam} \cite{Wheeler} is the intuitively correct
way to envision the emergence of quantum gravity, then the perspective taken in this paper must somehow
lead to low-level quantum fluctuations of space-time that have potentially observable consequences \cite{Ng}.
However, it must be stressed that an {\em arbitrary presumption} of such an effect may unintentionally serve
to mask uniquely identifiable quantum gravity signatures, simply by virtue of imposing a theoretical
prejudice {\em a priori}.
Without a sufficiently rigourous testing of the {\em existing physical theories}---when put under extreme
conditions---to induce an {\em observationally relevant} breakdown in the existing formalism,
the problem of disentangling and properly interpreting potential signatures of quantum gravity,
as opposed to more conventional explanations for the same phenomena, may introduce unnecessary challenges
towards discovering a successful theory of quantum gravity.

It seems evident, therefore, that having an acceptable notion of a {\em neighbourhood} of space-time is
crucial for achieving good progress towards identifying signatures of either quantum gravity, or at least
unanticipated physical effects that a successful quantum gravity theory has to incorporate.
This is because a quantum system involving relatively large Compton wavelengths will occupy some physical
extension in space-time that may reasonably be subjected to tidal forces generated by the curved space-time
background.
For example, it is recently shown that spin-1/2 particle {\em Zitterbewegung} in the
presence of a general curved space-time background leads to the prediction of weak equivalence principle violation \cite{Singh-Mobed3}.
Once other related issues are taken into proper account, a complete exploration of the overlap between $\lm$ and $l$ should
prove extremely useful for conceptualizing new physical phenomena to explain potentially unforeseen effects
that result from their intersection.

\subsection{Implications for Describing Covariant Spin Properties of Quantum Systems}

As it concerns interactions with the background gravitational field, intrinsic spin is a very useful tool
for probing the microscopic properties of space-time curvature.
The covariant formulation of spin angular momentum is given by the Pauli-Lubanski vector \cite{Itzykson}, whose
expression for spin-1/2 particles in flat space-time is given in the standard form
\be
\Wvec^\mu & = & -{1 \over 2} \, \varepsilon^\mu{}_{\al \bt \gm} \, \Jvec^{\al \bt} \, \Pvec^\gm \, ,
\label{Pauli-Lubanski-general}
\ee
where $\Pvec_\al = i \hb \, \pt{x^\al}$ is the momentum generator given in local Cartesian co-ordinates $x^\mu$ and
$\Jvec^{\al \bt}$ is the total angular momentum operator, also the infinitesimal generator of Lorentz transformations.
It becomes evident that, in the absence of space-time curvature and given that the momentum generators automatically
commute, the orbital angular momentum part of $\Jvec$ makes no contribution when $\Wvec$ is contracted with itself, resulting in
the Casimir invariant for spin,
\be
\Wvec \cdot \Wvec & = & -{1 \over 2} \lt({1 \over 2} + 1\rt) m^2 \, ,
\label{Casimir-spin}
\ee
where $m^2 = \Pvec \cdot \Pvec$.

This expression of Casimir invariance, as given in (\ref{Casimir-spin}) and within the context of localization issues
in space-time, leads to the following important question to consider.
How reliable is its applicability within the framework of {\em classical} space-time curvature,
let alone a comprehensive theory of quantum gravity?
It should be evident from the operator definitions implied in (\ref{Pauli-Lubanski-general}) that
a co-ordinate bias is already introduced that may not be satisfactory for more general considerations
involving the transport of quantum objects in space-time.
This is of particular relevance for assessing the range of validity for (\ref{Casimir-spin}),
since it presumes that $\Wvec \cdot \Wvec$, as defined to exist on a {\em locally} flat space-time background at a
{\em mathematical point} without any consideration for its {\em local neighbourhood},
will automatically obey the weak equivalence principle, a concept already suggested to have potential limitations
at the quantum mechanical level \cite{Singh-Mobed3}.

In fact, it is recently suggested that the mathematical conditions required to denote the existence
of ``Casimir invariants'' within the Poincar\'{e} group are no longer satisfied when considering
{\em arbitrary motion} of the quantum particle in space-time, and particularly so while
in the presence of classical external fields \cite{Singh-Mobed4}, where the breakdown of Casimir invariance is entirely due to the
presence of $\Rvec$.
The momentum generators, represented in curvilinear form, along with the spin generators still manage
to form a Lie algebra, since $\Rvec$ and its derivatives can always be expressed as functions of $\Pvec$.
Nonetheless, the structure of the associated Lie algebra becomes much more complicated to analyze,
and it is particularly unclear whether true ``Casimir invariants'' can be found for this new group that are applicable for
arbitrary motion in space-time.
When these considerations are further applied to the study of higher-order spinning particles besides spin-1/2,
it is also shown that acceleration-induced effects which cause the breakdown of standard Casimir invariance
lead to an identifiable distinction between fermions and bosons with interesting physical predictions,
such as the existence of a {\em maximal acceleration} initially proposed years ago
by Caianiello \cite{Singh-Mobed4,Caianiello}.
These and other related issues indicate that a much deeper study of the hypothesis of locality
may be required in a variety of physical contexts.

\section{Covariant Dirac Equation in Fermi Normal Co-ordinates}
\label{sec3}

Consider the covariant Dirac equation \cite{Singh-Mobed2}
\be
\lt[i \gm^\mu(X) \lt(\partial_\mu + i \, \Gamma_\mu(X) \rt) - m/\hb\rt]\psi(X) & = & 0 \,
\label{Dirac-eq}
\ee
for a spin-1/2 particle of mass $m$, where $X^\mu = \lt(T, X^j\rt)$ is the Fermi normal co-ordinate system defined in a local
neighbourhood about the spin-1/2 particle's worldline parametrized by
proper time $T$, where $X^j$ is the Cartesian spatial co-ordinate orthogonal to the worldline,
$\partial_\mu = {\partial \over \partial X^\mu}$, and $\Gamma_\mu(X)$ is the spin connection.
The gamma matrices $\lt\{ \gm^\mu(X) \rt\}$ satisfy $\lt\{ \gm^\mu(X), \gm^\nu(X) \rt\} = 2 \, g_F^{\mu \nu}(X)$,
where the space-time metric is described in terms of $\d s^2 = {}^F{}g_{\mu \nu}(X) \, \d X^\mu \, \d X^\nu$ by
%
\be
{}^F{}g_{00}(X) & = & 1 - {}^F{}R_{l00m}(T) \, X^l \, X^m + \cdots \, ,
\label{F-g00}
\nl
{}^F{}g_{0j}(X) & = & -{2 \over 3} \, {}^F{}R_{l0jm}(T) \, X^l \, X^m + \cdots \, ,
\label{F-g0j}
\nl
{}^F{}g_{ij}(X) & = & \eta_{ij} - {1 \over 3} \, {}^F{}R_{lijm}(T) \, X^l \, X^m + \cdots \, ,
\label{F-gij}
\ee
%
where $\eta_{\mu \nu}$ is the Minkowski metric, and ${}^F{}R_{\mu \al \bt \nu}(T)$ is the projection of the Riemann tensor onto
the Fermi frame.
The metric can be expressed in terms of orthonormal vierbeins $\lt\{ \bar{e}^\mu{}_{\hat{\alpha}}(X) \rt\}$
and inverse vierbeins $\lt\{ \bar{e}^{\hat{\alpha}}{}_\mu (X)\rt\} \,$ projected from a local Lorentz frame
defined by co-ordinates with hatted indices, such that
${}^F{}g_{\mu \nu}(X) = \eta_{\hat{\alpha}\hat{\beta}} \, \bar{e}^{\hat{\alpha}}{}_\mu (X) \, \bar{e}^{\hat{\beta}}{}_\nu (X)$.

It is straightforward to express (\ref{Dirac-eq}) in curvilinear co-ordinate form.
To begin, assume that $X^j = X^j(u^1, u^2, u^3)$, where $u^j$ is the curvilinear spatial co-ordinate.
Then the Fermi normal co-ordinate system takes the form $U^\mu$, where $U^0 = T$ and $U^j = u^j$.
A corresponding set of orthonormal vierbeins is then obtained, where
$e^\bt{}_{\hat{\al}}(U) = {\partial U^\bt \over \partial X^\al} \, \bar{e}^\al{}_{\hat{\al}}(X)$ and
$e^{\hat{\al}}{}_\bt(U) = {\partial X^\al \over \partial U^\bt} \, \bar{e}^{\hat{\al}}{}_\al(X)$.
Upon projecting onto the local Lorentz frame, (\ref{Dirac-eq}) in curvilinear co-ordinates is expressed as
\be
\lt[i \gm^{\hat{\mu}} \lt(\hat{\nabla}_{\hat{\mu}} + i \, \Gm_{\hat{\mu}}(U) \rt) - m/\hb\rt]\psi(U) & = & 0 \, ,
\label{Dirac-eq-curvilinear}
\ee
where $\hat{\nabla}_{\hat{\mu}} = \nabvec_{\hat{\mu}} + i \, \hat{\Gm}_{\hat{\mu}}(U)$ is the covariant derivative operator
defined with respect to Minkowski space-time in curvilinear co-ordinate form, with
$\nabvec_{\hat{0}} \equiv {\partial \over \partial T}$ and
$\nabvec_{\hat{\jmath}} \equiv {1 \over \lm^{(\hat{\jmath})}(u)} \, {\partial \over \partial u^{\hat{\jmath}}}$,
where $\lm^{(\hat{\mu})}(u)$ is the dimensional scale function and $\hat{\Gm}_{\hat{\mu}}(U)$ is the
corresponding spin connection.
From (\ref{Dirac-eq-curvilinear}), it follows that
\be
\lt[\gm^{\hat{\mu}} \lt(\Pvec_{\hat{\mu}} - \hb \, \Gmvec_{\hat{\mu}}(U) \rt) - m\rt]\psi(U) & = & 0 \, ,
\label{Dirac-eq-curvilinear-1}
\ee
where
\be
\Pvec_{\hat{\mu}} & = & \pvec_{\hat{\mu}} + \Omvec_{\hat{\mu}}
\label{Pvec}
\ee
is the momentum operator in curvilinear co-ordinates,
in terms of
%
\be
\pvec_{\hat{\mu}} & = & i \hbar \, \nabvec_{\hat{\mu}} \, ,
\label{pvec}
\nl
\Omvec_{\hat{\mu}} & = & i \hbar \lt[\nabvec_{\hat{\mu}} \ln \lt(\lm^{(\hat{1})}(u) \, \lm^{(\hat{2})}(u) \, \lm^{(\hat{3})}(u)\rt)^{1/2} \rt] \, .
\label{Ovec}
\label{Spin-Connection}
\ee
%
By now defining $\Dvec_{\hat{\mu}} = \Pvec_{\hat{\mu}} - \hb \, \Gmvec_{\hat{\mu}}$ and making use
of the identity \cite{Aitchison}
\be
\gm^{\hat{\mu}} \, \gm^{\hat{\nu}} \, \gm^{\hat{\rho}} & = & \eta^{\hat{\nu} \hat{\rho}} \, \gm^{\hat{\mu}}
- 2 \, \gm^{[\hat{\nu}} \eta^{\hat{\rho}]\hat{\mu}} - i \, \gm^5 \, \gm^{\hat{\sg}} \, \varepsilon^{\hat{\mu} \hat{\nu} \hat{\rho}}{}_{\hat{\sg}} \, ,
\label{gm-identity}
\ee
where $\varepsilon^{\hat{\mu} \hat{\nu} \hat{\rho} \hat{\sg}}$ is the Levi-Civita symbol with
$\varepsilon^{\hat{0} \hat{1} \hat{2} \hat{3}} = 1$, it is shown that
\be
\Dvec_{\hat{\mu}} & = & \Pvec_{\hat{\mu}} -
\hb \lt(\gm^5 \, \bar{\Gmvec}^{(\rm C)}_{\hat{\mu}} - i \, \bar{\Gmvec}^{(\rm S)}_{\hat{\mu}}\rt) \, ,
\label{D}
\nl
\bar{\Gmvec}^{(\rm C)}_{\hat{\mu}} & = & {1 \over 2} \, \varepsilon^{\hat{0} \hat{l} \hat{m}}{}_{\hat{\mu}} \,
{}^F{}R_{lm0k}(T) \, X^k  \, ,
\label{Spin-Connection-C}
\nl
\bar{\Gmvec}^{(\rm S)}_{\ze} & = & {1 \over 12} {}^F{}R^m{}_{jmk,0}(T) \, X^j \, X^k \, ,
\label{Spin-Connection-0-S}
\nl
\bar{\Gmvec}^{(\rm S)}_{\hat{\jmath}} & = & - {1 \over 2} {}^F{}R_{j00m}(T) \, X^m \, ,
\label{Spin-Connection-j-S}
\ee
where the ``C'' refers to the chiral-dependent part of the spin connection and the ``S'' denotes the symmetric part under chiral symmetry.

\section{Violation of Casimir Invariance for Spin-1/2 Particles}
\label{sec4}

Given the Pauli-Lubanski vector in the local Lorentz frame \cite{Singh-Mobed2}
\be
\Wvec^{\hat{\mu}} & = & -{1 \over 4} \, \varepsilon^{\hat{\mu}}{}_{\hat{\al}\hat{\bt}\hat{\gm}} \,
\sgvec^{\hat{\al}\hat{\bt}} \, \Dvec^{\hat{\gm}} \, ,
\label{Wvec-grav-em}
\ee
where $\sgvec^{\hat{\al}\hat{\bt}} = {i \over 2} \lt[\gm^{\hat{\al}} \, , \gm^{\hat{\bt}}\rt]$
and recognizing that the orbital angular momentum part $\Lvec^{\hat{\al}\hat{\bt}}$
of $\Jvec^{\hat{\al}\hat{\bt}} = \Lvec^{\hat{\al}\hat{\bt}} + {1 \over 2} \, \sgvec^{\hat{\al}\hat{\bt}}$
anticipated via (\ref{Pauli-Lubanski-general}) identically vanishes, it is straightforward to show that while
$m_0^2 = \Pvec^{\hat{\al}} \, \Pvec_{\hat{\al}}$ is a Lorentz invariant for momentum, the corresponding quantity for spin is not.
That is,
\be
\Wvec^{\hat{\al}} \, \Wvec_{\hat{\al}} & = & -{3 \over 4} \lt[m_0^2 + {\hbar^2 \over 6} \lt({}^F{}R^{\hat{\al}}{}_{\hat{\al}}\rt) \rt]
+ {\hbar \over 2} \lt(\sgvec \cdot \Rvec\rt)
\nn
&  &{} -{\hbar \over 4} \, \sg^{\hat{\al} \hat{\bt}} \, Q_{\hat{\al}\hat{\bt}}
+ {3 \over 2} \, \hbar \lt(\gm^5 \, \bar{\Gmvec}^{(\rm C)}_{\hat{\al}} - i \, \bar{\Gmvec}^{(\rm S)}_{\hat{\al}}\rt) \Pvec^{\hat{\al}}
\nn
&  &{} + {3 \over 4} \, \hbar^2 \, \nabvec^{\hat{\al}}
\lt(\bar{\Gmvec}^{(\rm S)}_{\hat{\al}} + i \, \gm^5 \, \bar{\Gmvec}^{(\rm C)}_{\hat{\al}}\rt) \,
\label{W^2-2}
\ee
in the presence of a curved space-time background, where
\be
{\hbar \over 2} \lt(\sgvec \cdot \Rvec\rt)
& = & - {\hbar \over 4} \, \sgvec^{\hat{\al} \hat{\bt}} \, \lt(\nabvec_{\hat{\al}} \, \ln \lm^{(\bt)}\rt) \Pvec_{\hat{\bt}} \,
\label{sg.R-defn}
\ee
in index notation---noting that the scale factors $\lm^{(\al)}(u)$ are already assumed to have no proper time
dependence and $\lm^{(0)}(u) = 1$---with
\be
\Rvec^{\hat{k}} & = & \lt. {i \over 2\hbar} \, \varepsilon^{\hat{0} \hat{\imath} \hat{\jmath} \hat{k}} \,
[\Pvec_{\hat{\imath}}, \Pvec_{\hat{\jmath}}] \rt|_{{}^F{}R_{\hat{\mu}\hat{\nu}\hat{\al}\hat{\bt}} \rightarrow 0}
\ = \ \varepsilon^{\hat{0} \hat{\imath} \hat{\jmath} \hat{k}} \,
\lt(\nabvec_{\hat{\imath}} \, \ln \lm^{(j)}\rt) \Pvec_{\hat{\jmath}} \,
\label{R-defn}
\ee
to denote the non-inertial dipole operator,
\be
Q_{\hat{\al}\hat{\bt}}
& = & 2 \lt(\bar{e}^\sg{}_{[\hat{\al}} \, \bar{e}^\gm{}_{\hat{\bt}]} - \dl^\sg{}_{[\hat{\al}} \, \dl^\gm{}_{\hat{\bt}]}\rt)
\lt(\nabvec_{\hat{\sg}} \, \ln \lm^{(\gm)}\rt) \Pvec_{\hat{\gm}}
+ C^{\hat{\mu}}{}_{\hat{\al}\hat{\bt}} \, \Pvec_{\hat{\mu}} \, ,
\label{Q-defn}
\ee
and
\be
C^{\hat{\mu}}{}_{\hat{\al}\hat{\bt}} & = & 
2 \, \bar{e}^{\hat{\mu}}{}_\sg \lt( \nabla_\lm \, \bar{e}^\sg{}_{[{\hat{\al}}} \rt) \bar{e}^\lm{}_{{\hat{\bt}}]} \,
\label{C-defn}
\ee
is the object of anholonomicity in the local Lorentz frame.
Both (\ref{Q-defn}) and (\ref{C-defn}) are exclusively first-order in ${}^F{}R_{\hat{\mu}\hat{\nu}\hat{\al}\hat{\bt}}$.
Clearly, (\ref{W^2-2}) reduces to its flat space-time counterpart in the absence of
external gravitational fields \cite{Singh-Mobed1}, with the non-inertial dipole interaction due to (\ref{R-defn}).
It is important to again emphasize that $\Pvec^{\hat{\al}} \, \Pvec_{\hat{\al}}$ and $\Wvec^{\hat{\al}} \, \Wvec_{\hat{\al}}$
are {\em not} the ``Casimir invariants'' for the Poincar\'{e} group in curvilinear co-ordinates,
since neither operator commutes with all the generators of the Lie algebra associated with the Poincar\'{e} group,
a defining property of Casimir invariance \cite{Singh-Mobed4}.

\section{Muon Decay Near a Microscopic Kerr Black Hole}
\label{sec5}

While the expression (\ref{W^2-2}) is interesting by itself, a fuller appreciation comes when the Pauli-Lubanski vector
is applied to a specific physical process, whose spectrum may reveal interesting observations
worth exploring in greater detail.
This is evident for the example of muon decay while orbiting a microscopic Kerr black hole in the equatorial plane
near its event horizon.
Consider the reaction $\mu^- \rightarrow e^- + \bar{\nu}_e + \nu_\mu$ in a gravitational background.
Then the associated matrix element is \cite{Singh-Mobed1,Singh-Mobed2}
\be
|{\cal M}|^2 & = &
{G_{\rm F}^2 \over 2} \, L^{(\mu)}_{\hat{\mu} \hat{\nu}} \, M^{\hat{\mu} \hat{\nu}}_{(e)},
\label{M^2}
\nl
L^{\hat{\mu} \hat{\nu}}_{(\mu)} & = & {\rm Tr} \lt[\gslash{\pvec}_{\nu_\mu} \, \gm^{\hat{\mu}} \lt(\gslash{\Dvec}_{\mu}
+ m_{\mu} \, \gfive \, \gslash{\nvec}_{\mu}\rt)\gm^{\hat{\nu}} \lt(1 - \gfive\rt)\rt] \, ,
\label{L_uv}
\nl
M^{\hat{\mu} \hat{\nu}}_{(e)} & = & {\rm Tr} \lt[\lt(\gslash{\Dvec}_{e} + m_{e} \, \gfive \, \gslash{\nvec}_e\rt)
\gm^{\hat{\mu}} \, \gslash{\pvec}_{\nu_e} \, \gm^{\hat{\nu}} \lt(1 - \gfive\rt)\rt] \, ,
\label{M_uv}
\ee
%
where $\nvec^{\hat{\mu}}$ is the polarization vector for the charged lepton,
satisfying $\nvec^{\hat{\mu}} \, \Dvec_{\hat{\mu}} = 0$.
In flat space-time, the polarization vector is $\nvec^{\rm (0)}_{\hat{\mu}} = {E \over m_0 |\Pvec|} \, \Pvec_{\hat{\mu}} - {m_0 \over |\Pvec|}
 \, \dl^{\hat{0}}{}_{\hat{\mu}}$, where $E = |\Pvec^{\hat{0}}|$.
Though tedious, it is straightforward to show that
$\gslash{\nvec} = \lt[\gslash{\nvec^{\rm (S, Re)}} + i \, \gslash{\nvec^{\rm (S, Im)}}\rt]
+ \gm^5 \lt[\gslash{\nvec^{\rm (C, Re)}} + i \, \gslash{\nvec^{\rm (C, Im)}}\rt]$, where
\be
\nvec^{\rm (S, Re)}_{\hat{\mu}} & \approx & \nvec^{\rm (0)}_{\hat{\mu}}
+ {4 \hbar \, E \over m_0^3 |\Pvec|} \, \dl^{\hat{0}}{}_{[\hat{\mu}} \, \dl^{\hat{\jmath}}{}_{\hat{\nu}]} \, \Rvec_{\hat{\jmath}} \, \Pvec^{\hat{\nu}}
- {\hbar^2 \, E \over m_0^3 |\Pvec|} \lt(\nabvec^{\hat{\nu}} Q_{\hat{\mu} \hat{\nu}}\rt) \, ,
\label{n-S,Re}
\nl \nn \nn
\nvec^{\rm (S, Im)}_{\hat{\mu}} & \approx &
{2 \hbar^2 \, E \over m_0^3 |\Pvec|} \, \dl^{\hat{0}}{}_{[\hat{\mu}} \, \dl^{\hat{\jmath}}{}_{\hat{\nu}]}
\lt(\nabvec^{\hat{\nu}} \Rvec_{\hat{\jmath}}\rt) + {\hbar \, E \over m_0 |\Pvec|} \, \bar{\Gmvec}^{\rm (S)}_{\hat{\mu}}
\nn
&  &{} - {4 \hbar \over m_0 |\Pvec|} \lt[\dl^{\hat{0}}{}_{\hat{\mu}}
\lt(\bar{\Gmvec}_{\rm (S)}^{\hat{\lm}} \, \Pvec_{\hat{\lm}} \rt)
+ {1 \over 2} \, \varepsilon^{\hat{0} \hat{\al} \hat{\bt}}{}_{\hat{\mu}} \, \bar{\Gmvec}^{\rm (C)}_{\hat{\al}} \, \Pvec_{\hat{\bt}} \rt]
\nn
&  &{} - {2 \hbar \over m_0^2} \, \nvec^{\rm (0)}_{\hat{\nu}} \lt[\dl^{\hat{\nu}}{}_{\hat{\mu}}
\lt(\bar{\Gmvec}_{\rm (S)}^{\hat{\lm}} \, \Pvec_{\hat{\lm}} \rt)
+ \varepsilon^{\hat{\nu} \hat{\al} \hat{\bt}}{}_{\hat{\mu}} \, \bar{\Gmvec}^{\rm (C)}_{\hat{\al}} \, \Pvec_{\hat{\bt}} \rt]
\nn
&  &{} + {\hbar \, E \over m_0^3 |\Pvec|}\lt[
2 \lt(\bar{\Gmvec}_{(\rm S)}^{\hat{\lm}} \, \Pvec_{\hat{\lm}} \rt) \Pvec_{\hat{\mu}}
- {\hbar^2 \over 4} \, \nabvec_{\hat{\mu}} \lt({}^F{}R^{\hat{\al}}{}_{\hat{\al}}\rt) \rt]
+ {2 \hbar \, E \over m_0^3 |\Pvec|} \,  Q_{\hat{\mu} \hat{\nu}} \, \Pvec^{\hat{\nu}} \, ,
\label{n-S,Im}
\nl \nn \nn
\nvec^{\rm (C, Re)}_{\hat{\mu}} & \approx &
-{\hbar^2 \, E \over m_0^3 |\Pvec|} \, \varepsilon^{\hat{\sg} \hat{\al} \hat{\bt}}{}_{\hat{\mu}} \,
\dl^{\hat{0}}{}_{[\hat{\al}} \, \dl^{\hat{\jmath}}{}_{\hat{\bt}]}
\lt(\nabvec_{\hat{\sg}} \Rvec_{\hat{\jmath}}\rt)
\nn
&  &{} + {3 \hbar \, E \over m_0 |\Pvec|} \, \bar{\Gmvec}^{\rm (C)}_{\hat{\mu}}
- {2 \hbar \over m_0 |\Pvec|} \, \dl^{\hat{0}}{}_{\hat{\mu}} \, \lt(\bar{\Gmvec}_{\rm (C)}^{\hat{\lm}} \, \Pvec_{\hat{\lm}} \rt)
\nn
&  &{} - {2 \hbar \, E \over m_0^3 |\Pvec|}\lt[
\eta_{\hat{\mu} \hat{\nu}} \lt(\bar{\Gmvec}_{\rm (S)}^{\hat{\lm}} \, \Pvec_{\hat{\lm}} \rt)
+ 2 \, \bar{\Gmvec}^{\rm (C)}_{[\hat{\mu}} \, \Pvec_{\hat{\nu}]} \rt] \Pvec^{\hat{\nu}} \, ,
\label{n-C,Re}
\nl \nn \nn
\nvec^{\rm (C, Im)}_{\hat{\mu}} & \approx &
-{\hbar^2 \, E \over 2 m_0^3 |\Pvec|} \,
\varepsilon^{\hat{\sg} \hat{\al} \hat{\bt}}{}_{\hat{\mu}} \lt(\nabvec_{\hat{\sg}} Q_{\hat{\al} \hat{\bt}}\rt) \, ,
\label{n-C,Im}
\ee
to leading order in curvature.
Substitution of (\ref{L_uv})--(\ref{n-C,Im}) into (\ref{M^2}) leads to
$|{\cal M}|^2 = 32 \, G_{\rm F}^2 \lt(\Pvec_{\nu_e}^{\hat{\al}} \, \bar{\Dvec}^{\mu}_{\hat{\al}} \rt)
\lt(\Pvec_{\nu_\mu}^{\hat{\bt}} \, \bar{\Dvec}^{e}_{\hat{\bt}} \rt)$,
where
$\bar{\Dvec}_{\hat{\al}} = \Pvec_{\hat{\al}} - \hbar \lt(\bar{\Gmvec}^{\rm (C)}_{\hat{\al}} - i \, \bar{\Gmvec}^{\rm (S)}_{\hat{\al}} \rt)
+ m_0 \lt(\nvec_{\hat{\al}}^{\rm (C)} - \nvec_{\hat{\al}}^{\rm (S)}\rt)$.

Formally, the gravitational and $\Rvec$-dependent contributions to the muon decay rate are additive corrections to
$\Gamma_0 \approx G_{\rm F}^2 \, m_\mu^5/(192 \, \pi^3) \approx 2.965 \times 10^{-16}$~MeV \cite{Singh-Mobed1}.
This becomes evident when applied to circular motion around a microscopic Kerr black hole
of mass $M = 2 \times 10^{-11}$~cm \cite{Singh-Mobed2},
about an order of magnitude larger than the muon's Compton wavelength of $1.18 \times 10^{-12}$~cm.
Suppose that the black hole's spin angular momentum is $a$ with $-M \leq a \leq M$ and $\al \equiv a/r_0$, where $r_0$
is the radius of the muon's orbit prior to decay.
Then for the innermost (photon) orbits of $r_{0+} = M$ and $r_{0-} = 4M$ for co-rotating and counter-rotating black holes,
respectively, it follows that $-1/4 \leq \al \leq 1$ and
\be
r_0 & = & 9M \lt[\al + \sqrt{\al^2 + 3 \lt(1 - N_0^2\rt)} \rt]^{-2} \, ,
\label{r0}
\ee
where $N_0 \geq 0$ is the separation away from the innermost circular orbit.

\begin{figure}
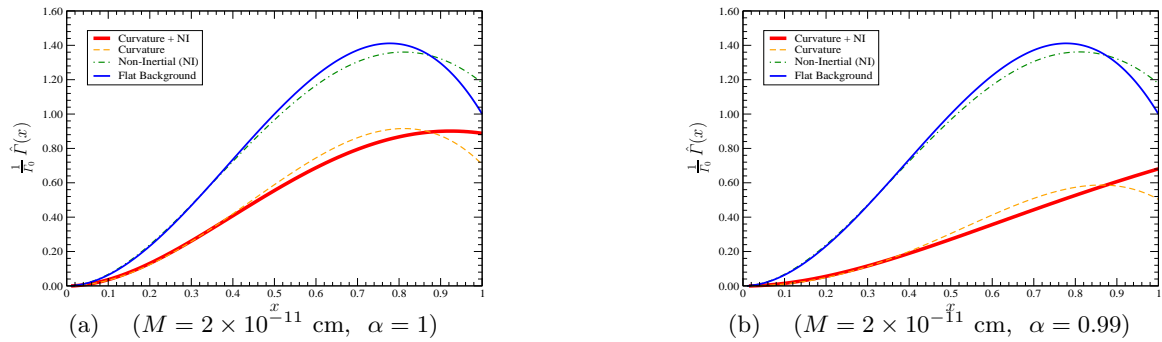

\psfrag{x}[tc][][1.8][0]{\Large $x$}
\psfrag{dG}[bc][][1.8][0]{\Large ${1 \over \Gm_0} \, \hat{\Gm}(x)$}
\vspace{4mm}
\begin{minipage}[t]{0.3 \textwidth}
\centering
\subfigure[\hspace{0.2cm} ($M = 2 \times 10^{-11}$ cm, \ $\al = 1$)]{
\label{fig:Michel-r0=2e-11-al=100}
\rotatebox{0}{\includegraphics[width = 6.0cm, height = 4.0cm, scale = 1]{Michel-M=2e-11-N=1e-2-alpha=+100}}}
\end{minipage}
\hspace{4cm}
\begin{minipage}[t]{0.3 \textwidth}
\centering
\subfigure[\hspace{0.2cm} ($M = 2 \times 10^{-11}$ cm, \ $\al = 0.99$)]{
\label{fig:Michel-r0=2e-11-al=099}
\rotatebox{0}{\includegraphics[width = 6.0cm, height = 4.0cm, scale = 1]{Michel-M=2e-11-N=1e-2-alpha=+099}}}
\end{minipage}
\caption{\label{fig:Michel-r0=2e-11} Michel spectrum with contributions due to curvature and $\Rvec$
for $M = 2 \times 10^{-11}$ cm and $r_0 \sim M$.
It is clear from comparing Figs.~\ref{fig:Michel-r0=2e-11-al=100} and \ref{fig:Michel-r0=2e-11-al=099}
that both curvature variations and non-inertial effects can strongly influence the shape of the plots
in relation to the flat space-time contribution.
}
\end{figure}
Computing in terms of the muon's local rest frame and assuming $N_0 = 10^{-2}$,
Figure~\ref{fig:Michel-r0=2e-11} displays the Michel spectrum of the muon decay $\hat{\Gm}(x) = \d \Gm/\d x$
as a function of the outgoing electron's energy fraction $x$ and in units of $\Gm_0$.
Comparison between Fig.~\ref{fig:Michel-r0=2e-11-al=100} and Fig.~\ref{fig:Michel-r0=2e-11-al=099}
clearly shows that a 1\% reduction in $\al$ from the extremal black hole condition $\al = 1 \ (r_0 \sim M)$
reveals an extreme sensitivity of the Michel spectrum due to small variations in curvature.
It is self-evident that, in the absence of curvature, the non-inertial dipole operator $\Rvec$
introduces a noticeable correction to the flat space-time Michel spectrum, indicating a prediction
that may be potentially observable via precision measurements of muon decay for large values of $x$
under suitable experimental conditions.
As well, the curvature contribution to the muon decay spectrum shows a reduction of the decay rate that
grows smaller as $r_0 \rightarrow 0$, indicating stabilization of the muon in the absence of unforeseen
physical effects due to a truly quantum mechanical theory of gravity.

\section{Conclusion}

This paper shows that an exploration of the Pauli-Lubanski vector for spin-1/2 particles in general
motion while in a curved space-time background leads to the prediction of Lorentz invariance violation at the
Compton wavelength scale, with interesting and potentially observable consequences for muon decay.
The result presented here is consistent with other evidence \cite{Rosquist} of emergent and previously unanticipated phenomena
occurring at the Compton wavelength scale due to this combination of general relativity and quantum mechanics.
Though not discussed in this paper, it is also possible to identify theoretical signatures of noncommutative geometry
in the Michel spectrum \cite{Singh-Mobed2,Singh-Mobed-Ouimet},
which appear for $x > 0.65$ in the specific cases considered \cite{Singh-Mobed-Ouimet}
when combined with Moffat's theory of nonsymmetric gravity \cite{Moffat1,Moffat2}.

If more fundamental approaches towards quantum gravity are to be deemed viable, they will need to properly
account for this type of phenomenon, should it become observable in the near future.
This possibility is strongly dependent on first confirming the existence of microscopic black holes,
followed by precision observations of their behaviour in the presence of quantum matter in order to
positively identify muon decay suppression near the event horizon.
At present, this seems confined to the realm of conceptual interest only.
However, precision measurements involving muonic atoms \cite{Singh-Mobed2} in a flat background
offer a more promising capacity to observe $\Rvec$-dependent contributions to the Michel spectrum.
Of course, this requires a much more detailed computational analysis in the presence of an effective
potential due to the muonic atom's atomic nucleus, but this should be within the realm of possibility
in the future.

One of the prevailing conclusions drawn from this approach is the claim that
$\Pvec^{\hat{\al}} \, \Pvec_{\hat{\al}}$ and $\Wvec^{\hat{\al}} \, \Wvec_{\hat{\al}}$ can no longer be faithfully
described as ``Casimir invariants'' for the Poincar\'{e} group if the generators $\Jvec^{\hat{\al}\hat{\bt}}$ and $\Pvec^{\hat{\gm}}$
are expressed in terms of curvilinear co-ordinates.
The interpretation of this outcome offered by this paper is the prediction of an emergent frame-induced violation of
Lorentz invariance, stemming from limitations with the hypothesis of locality.
However, it is possible to instead conclude that the new effects predicted are merely mathematical constructs
resulting from a particular {\em representation} of the Lie algebra generators for the Poincar\'{e} group.
In order to ascertain the merits of this other possibility, it would be necessary to explore the status
of $\Jvec^{\hat{\al}\hat{\bt}}$ and $\Pvec^{\hat{\gm}}$ within the framework of the universal enveloping algebra
for the Poincar\'{e} group, and determine whether the effects predicted in this paper are indeed physical in origin or
merely the byproducts of a particular representation.
This is an open question worthy of deeper exploration.




\end{document}